\documentclass[aps,epsfig,twocolumn]{revtex4}
\usepackage{amsmath}
\usepackage{epsfig}
\usepackage{graphics}
\begin{document}
\title{Detection of BCS pairing in neutral Fermi fluids via
Stokes scattering:\\ the Hebel-Slichter effect}
\author{G.\ M.\ Bruun$^1$ and Gordon\ Baym$^{2,3}$}
\affiliation{$^1$Niels Bohr Institute, Blegdamsvej 17, 2100 Copenhagen,
Denmark\\
$^2$NORDITA, Blegdamsvej 17, 2100 Copenhagen, Denmark\\
$^3$Department of Physics, University of Illinois, 1110 W.\ Green St.,
Urbana, IL 61801}

\begin{abstract}
    We consider the effects of superfluidity on the light scattering
properties of a two component gas of fermionic atoms, demonstrating that the
scattered intensities of the Stokes and anti-Stokes lines exhibit a large
maximum below the critical temperature when the gas is superfluid.  This
effect, the light scattering analogue of the Hebel-Slichter effect in
conventional superconductors, can be used to detect unambiguously the onset of
superfluidity in an atomic gas in the BCS regime.
\end{abstract}

\maketitle

Pacs Numbers: 03.75.Ss 05.30.Fk 32.80.-t\

    Nearly a decade after the first experimental reports of atomic
Bose-Einstein condensates (BEC)~\cite{BEC}, the trapping and cooling of gases
with Fermi statistics has become one of the central areas of research within
the field of ultracold atomic gases.  Such gases offer the exciting prospect
of examining the properties of interacting Fermi gases, including BCS
superfluid states, with unprecedented accuracy and flexibility.

    Using Feshbach resonances, one can readily vary the interactions between
atoms, allowing study of the crossover between a BEC of tightly bound
molecules on the side of the resonance where $0<k_Fa\ll 1$, with $a$ the
scattering length and $k_F$ the Fermi momentum, to a BCS superfluid state on
the other side of the resonance where $0<-k_Fa\ll1$, with an interesting
crossover regime in between~\cite{Holland,Griffin}.  Indeed, several groups
have now reported clear experimental results for a BEC of diatomic molecules
by looking at the momentum distribution of the gas~\cite{BECmol}.  Experiments
have also probed the $a<0$ side of the resonance using a fast magnetic sweep
to the BEC side~\cite{BCSside}.  Very recently, the $a<0$ side ($-k_Fa\gtrsim
3.3$) was
probed using a radio frequency (rf) transition to an unpaired hyperfine
state~\cite{Chin}.  The observed broadening of the rf spectral line was
interpreted as arising from formation of Cooper pairs~\cite{Kinnunen}.  For a
trapped system this effect is complicated by the fact that the mean (Hartree)
field also yields a significant broadening of the spectrum; the unpaired
hyperfine state in general sees a different spatially dependent Hartree field
than the paired ones, giving non-trivial energy shifts in the hyperfine
transitions.  The broadening of the rf line due to this effect has been
observed experimentally \cite{Gupta}, and it has been demonstrated for a
spherical gas in the BCS regime that it introduces broadening of the rf
spectrum at least comparable to the effects coming from
pairing~\cite{BruunLaser}.  An interpretation of the broadening of the rf
spectrum in terms of pairing effects only is therefore not straightforward for
an inhomogeneous system.  The absence of unambiguous signatures of the
presence of superfluidity in the BCS limit ($k_F|a|\ll 1$), arises because the
formation of large delocalized Cooper pairs does not affect the bulk
properties.  Clear observation of BCS superfluidity is one of the central
problems for experimentalists.  Suggestions to detect the onset of
superfluidity include probing the collective mode spectrum~\cite{BruunMott},
the quantization of angular momentum~\cite{Vortex}, off-resonance light
scattering~\cite{Zhang,Weig}, and by probing the dynamic structure factor
using a scheme which avoids the complication due to the inhomogeneous Hartree
field \cite{Laserprobing}.

\begin{figure}[htbp]
\begin{center}\vspace{0.0cm}
\rotatebox{0}{\hspace{-0.cm}
\resizebox{7.5cm}{!}
{\includegraphics{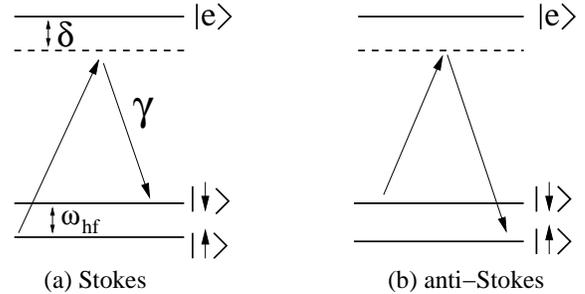}}}
    \caption{Stokes (a) and anti-Stokes (b) scattering of an incident laser
beam.  In (a) the laser, detuned off resonance by a frequency $\delta$,
excites a particle in the lower level
$|\uparrow\rangle$ to an intermediate highly excited level $|\uparrow\rangle$,
which de-excites, emitting a lower frequency photon, $\gamma$.  The hyperfine
splitting of the paired levels is $\omega_{\rm hf}$.  In (b) the initial and
final states are interchanged.
    }
\label{StokeFig}
\end{center}
\end{figure}

    One of the hallmark experimental tests of BCS theory in conventional
superconductors was the observed enhancement of the nuclear spin relaxation
just below the transition temperature \cite{Hebel}, an experiment which probed
the detailed nature of the pair correlations.  In this paper, we propose a
related experiment to probe the pair correlations in trapped atomic gases, by
looking at off-resonance inelastic (Stokes and anti-Stokes) light scattering
of a laser beam on a two component atomic Fermi gas.  The present scheme
differs from the earlier suggestions using off-resonant light
scattering~\cite{Zhang,Weig} crucially:  we propose looking at Stokes and
anti-Stokes inelastic light scattering involving a change of the hyperfine
states of the atoms.  The present proposal has several attractive features.
First, there is no central coherent scattering peak, which is effectively
insensitive to superfluid correlations.  Also, as we show, the effects due to
superfluidity on the scattered spectrum can be strikingly large close to the
critical temperature, $T_c$.  Finally, since the present scheme simply flips
the atoms between the two paired hyperfine states, it is not beset by
complications due to non-trivial energy shifts coming from non-superfluid
effects, e.g., shifts in the Hartree energies \cite{Laserprobing,Kinnunen}.

    We consider a trapped gas of atoms in two equally populated hyperfine
states labelled by $|\sigma\rangle$, with $\sigma=\uparrow,\downarrow$,
interacting via an effective attractive interaction.  We take the state
$|\downarrow\rangle$ to have energy $\omega_{\rm hf}$ ($\hbar=1$ in this
paper) above the state $|\uparrow\rangle$, and assume that below a transition
temperature, $T_c$, the gas is BCS paired and superfluid.  Consider a laser
beam of frequency $\omega_L$ and wave vector $k$ illuminating the gas.  As
illustrated in Fig. 1, the light field can induce dipole radiation from the
atoms by connecting the two hyperfine states $\sigma=\uparrow,\downarrow$ to
an electronically excited state $|e\rangle$, which we take to have energy
$\omega_e$ above the state $|\uparrow\rangle$.

    For large laser detuning, $\delta=\omega_L-\omega_e$, the excited level
$|e\rangle$ is not significantly populated and it can be adiabatically
eliminated from the theory.  The spectral intensity of the scattered light at
position ${\mathbf{r}}$ is then
\begin{gather}
  S({\mathbf{r}},\omega)=\sum_{^{\sigma_1\sigma_2}_{\sigma_3\sigma_4}}
  I_{\sigma_2\sigma_3}^{\sigma_4\sigma_3}(\mathbf{r})\int_{-\infty}^\infty
  dt\int
  d^3r_1d^3r_2
 e^{i[\omega t+\Delta{\mathbf{k}}\cdot({\mathbf{r}}_1-{\mathbf{r}}_2)]}
\nonumber\\
  \times\langle
\psi^\dagger_{\sigma_4}({\mathbf{r}}_10)\psi_{\sigma_3}({\mathbf{r}}_10)
\psi^\dagger_{\sigma_2}({\mathbf{r}}_2t)\psi_{\sigma_1}({\mathbf{r}}_2t)\rangle
  \label{Mastereqn}
\end{gather}
where the $\psi_{\sigma}$ are the field operators for the two low-lying
hyperfine states $|\sigma\rangle$ \cite{Javanainen}.  Here
$I_{\sigma_2\sigma_3}^{\sigma_4\sigma_3}(\mathbf{r})$ is the scattered light
intensity from a single atom, including the dependence of the atomic levels
involved and the various directions of the experiment, and
$\Delta{\mathbf{k}}$ is the change in momentum of the scattered light compared
to the incident light.  The frequency $\omega=\omega_S-\omega_L$ is the
difference between the scattered light frequency $\omega_S$ and the incident
light frequency $\omega_{\rm L}$.

    Equation (\ref{Mastereqn}) describes two types of off-resonant light
scattering processes.  The first, elastic or coherent scattering is characterized by
the initial and final atomic hyperfine states being identical, corresponding
to $\sigma_4=\sigma_3$ and $\sigma_2=\sigma_1$.  This process has been
examined in detail by a number of authors and several effects of pairing
have been identified.  However, measurement of these effects is complicated by
a large background coherent scattering intensity which is largely independent
of the state of the gas~\cite{Zhang,Weig}.

    The second type of scattering process is characterized by different
initial and final atomic hyperfine states.  If the initial state of the atom
is $|\uparrow\rangle$ and the final is $|\downarrow\rangle$, the emitted light
frequency is reduced by $\omega_{\rm hf}$ from the incoming light
frequency; this Stokes scattering process, Fig.\ 1a,
corresponds to $\sigma_1=\sigma_4=\uparrow$ and
$\sigma_2=\sigma_3=\downarrow$ in Eq.\ (\ref{Mastereqn}).  If, on the other
hand, the initial state of the atom is $|\downarrow\rangle$ and the final is
$|\uparrow\rangle$ [$\sigma_1=\sigma_4=\downarrow$ and
$\sigma_2=\sigma_3=\uparrow$ in Eq.\ (\ref{Mastereqn})], the emitted light
frequency is increased by $\omega_{\rm hf}$; this anti-Stokes scattering
process is shown in Fig.\ 1b.  Stokes and anti-Stokes
scattering therefore probe correlation functions that involve a hyperfine
state ``spin" flip [at positions ${\mathbf{r}}_1$ and ${\mathbf{r}}_2$ in
Eq.(\ref{Mastereqn})].  As we show, the rate of these Stokes and anti-Stokes
transitions involving atomic hyperfine state flips are strongly affected close
to $T_c$ by BCS pairing of the states, in contrast to elastic light scattering
which does not involve a hyperfine flip.  We note that for such inelastic
scattering to occur, dipole transitions between the electronically excited
(orbital p) state $|e\rangle$ and the two (orbital s) hyperfine states should
be allowed by the selection rules.  That is, both the dipole matrix elements
relevant for Stokes and anti-Stokes scattering, $\langle
e|{\mathbf{d}}\cdot{\cal E}|\uparrow\rangle$ and $\langle
e|{\mathbf{d}}\cdot{\cal E}|\downarrow\rangle$, with ${\mathbf{d}}$ the atomic
dipole operator, and ${\cal E}$ the electric field of the incident laser, must
be non-zero.

    From Eq.\ (\ref{Mastereqn}), the problem of calculating the scattered
light intensity is reduced to evaluating the Fourier transform of the
correlation function $
\langle\psi^\dagger_{\sigma_4}({\mathbf{r}}_10)\psi_{\sigma_3}({\mathbf{r}}_10)
\psi^\dagger_{\sigma_2}({\mathbf{r}}_2t)\psi_{\sigma_1}({\mathbf{r}}_2t)
\rangle$.  Experiments on the pairing transition for ultracold Fermi atomic
gases use a Feshbach resonance to enhance the atom-atom interaction, thereby
increasing $T_c$.  The gas is then best regarded as a molecular BEC on one
side of the resonance and a weakly coupled BCS superfluid on the other side of
the resonance, with a crossover region in between~\cite{Holland,Griffin}.  We
are interested here in the problem of detecting the presence of large
delocalized Cooper pairs on the BCS side of the resonance, where it is
adequate to use mean-field BCS theory.  We expand the field operators in
Bogoliubov eigenmodes $[u_\eta({\mathbf{r}}),v_\eta({\mathbf{r}})]$ with
energy $E_\eta$, which can be obtained from a solution of the Bogoliubov-de
Gennes equations~\cite{deGennes}.  With this expansion, Eq.\
(\ref{Mastereqn}) for Stokes and anti-Stokes scattering yields,
\begin{gather}
 S(\omega)\propto\sum_{\eta\eta'} \Big\{
 \frac{1}{2}|(u_\eta^* v_{\eta'}^*-v_\eta^*
 u_{\eta'}^*)_{\Delta\mathbf{k}}|^2\nonumber\\
\times(1-f_\eta)(1-f_{\eta'})
\delta(\tilde{\omega}+E_{\eta'}+E_{\eta})\nonumber\\
+\frac{1}{2}|(u_\eta v_{\eta'}-v_\eta u_{\eta'})_{\Delta\mathbf{k}}|^2
f_\eta f_{\eta'}
\delta(\tilde{\omega}-E_{\eta'}-E_{\eta}) \nonumber\\
+|(u_\eta u_{\eta'}^*+v_\eta v_{\eta'}^*)_{\Delta\mathbf{k}}|^2
f_\eta(1-f_{\eta'})
\delta(\tilde{\omega}+E_{\eta'}-E_{\eta})\Big\},
\label{Stokes}
\end{gather}
where $f_\eta=[\exp(\beta E_\eta)+1]^{-1}$ is the Fermi function and
$h_{\mathbf{k}}=\int d^3r\exp(-i{\mathbf{kr}})h({\mathbf{r}})$ denotes the
Fourier transform.  Equation (\ref{Stokes}) yields the intensity of the Stokes
line if $\tilde{\omega}=\omega+\omega_{\rm hf}$ and that of the anti-Stokes
line if $\tilde{\omega}=\omega-\omega_{\rm hf}$.

    In Eq.\ (\ref{Stokes}), we have, for notational simplicity, omitted the
prefactor in $S(\omega)$ involving the atomic dipole matrix elements and the
directional dependences of the experimental setup.  The first two terms in
Eq.\ (\ref{Stokes}) describe the creation and annihilation, respectively, of
two quasiparticles.  The third term describes the scattering and hyperfine
"spin" flip of a quasiparticle.  For Stokes scattering, the quasiparticle is
in the initial state $(\eta\uparrow)$ with energy $E_\eta$ and it scatters
into the state $(\eta'\downarrow)$ with energy $E_\eta'+\omega_{\rm hf}$.
Energy conservation for this process reads
$\omega_L+E_\eta=\omega_S+E_\eta'+\omega_{\rm hf}$.  For anti-Stokes
scattering the initial and final quasiparticle states are $(\eta\downarrow)$
and $(\eta'\uparrow)$ with energies $E_\eta+\omega_{\rm hf}$ and $E_\eta'$
respectively.

    We first consider light scattering on a homogeneous system, which can be
approximately realized experimentally for a trapped gas by focusing the laser
beam on an area much smaller than the size of the atomic cloud.  The
quasiparticle eigenfunctions are then plane waves; for small frequency
shifts, $\tilde{\omega}\ll \Delta$, where $\Delta$ is the superfluid gap, we
can neglect the terms in Eq.\ (\ref{Stokes}) describing the creation and
annihilation of two quasiparticles, and obtain
\begin{gather}
S(\omega)\propto\sum_{\mathbf{q}}
 |u_q u_{{\mathbf{q}}+\Delta\mathbf{k}}+v_q
 v_{{\mathbf{q}}+\Delta\mathbf{k}}|^2
 f_q(1-f_{{\mathbf{q}}+\Delta\mathbf{k}})\nonumber\\
 \times\delta(\omega+\omega_{\rm
  hf}+E_{{\mathbf{q}}+\Delta\mathbf{k}}-E_q),
\label{Slichterk}
\end{gather}
with $u_q^2=(1+\xi_q/E_q)/2$ and $v_q^2=(1-\xi_q/E_q)/2$.  Here
$\xi_q=q^2/2m-\mu$, and $E_q=\sqrt{\xi_q^2+\Delta^2}$.  In the limit, $\Delta
k\ll k_F$, and $\tilde{\omega}\ll \Delta k k_F/m$ this expression reduces to,
\begin{gather}
  S(\omega)\propto\frac{1}{\Delta
  k}\int_{\rm{min}(\Delta,\Delta+\tilde{\omega})}^\infty dE
\frac{E}{\sqrt{E^2-\Delta^2}}\frac{E'}{\sqrt{E'^2-\Delta^2}}\nonumber\\
\left(1+\frac{\Delta^2}{EE'}\right)f(1-f'),
\label{Slichter}
\end{gather}
where $E'=E-\tilde{\omega}$.  Equation (\ref{Slichter}) is identical to
the expression for the nuclear relaxation rate for polarized nuclei in a
conventional superconductor~\cite{Schrieffer}.  In particular, we note that
the scattered intensity is logarithmically divergent for $\tilde{\omega}=0$ in
a infinite size superfluid system.  This apparent divergence, which arises
from the divergent density of states at the Fermi surface for a
superconductor, only appears in ``spin"-flip processes such as
Stokes and anti-Stokes scattering.  In reality the divergence is mediated by
finite size and finite lifetime effects.

    Had we instead considered elastic scattering processes where the initial
and final hyperfine states of the atom are identical,
the coherence factor in the last term in Eq.\
(\ref{Stokes}) describing quasiparticle scattering would read $uu'^*-vv'^*$.
This factor vanishes at the Fermi surface for $\tilde{\omega}=0$, and the
coherence term for elastic scattering exactly cancels the divergent density of
states at the Fermi surface.  For Stokes and anti-Stokes scattering on the
other hand, the coherence factor in Eq.\ (\ref{Stokes}) reads $uu'^*+vv'^*\sim
1$ and the divergence in the density of states at the Fermi surface due to
pairing shows up in the response of the gas.  The coherence factor reflects
the fact that the coupling of the quasiparticles to a ``spin flip"
perturbation is essentially the same in the superfluid and normal phases.
It is important that the measurement be carried out a finite $\Delta k$
such that $\tilde{\omega}\ll \Delta k k_F/m$.  For $\Delta k=0$, the overlap
integrals in Eq.\ (\ref{Stokes}) simply yield a $\delta_{\eta,\eta'}$
selection rule, and the scattered signal becomes $\propto \sum_\eta
f_\eta(1-f_\eta)$, which does not exhibit a peak below $T_c$ due to pairing.
Measuring the intensity of the Stokes and anti-Stokes lines allows one to test
the detailed nature of the pairing correlations reflected in the coherence
factors, as was done for the BCS theory \cite{Hebel}.

\begin{figure}[htbp]
\begin{center}\vspace{0.0cm}
\rotatebox{0}{\hspace{-0.cm}
\resizebox{7.5cm}{!}
{\includegraphics{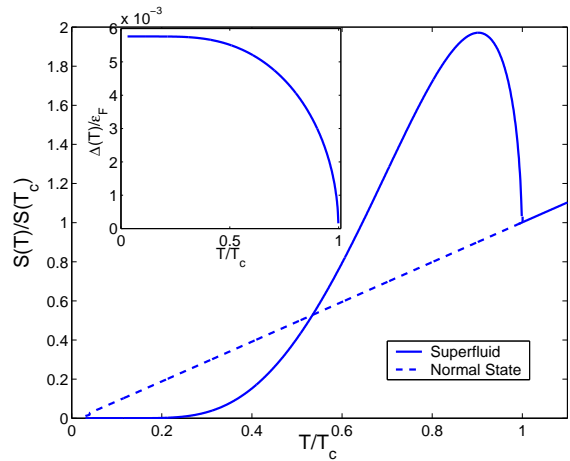}}}
\caption{The scattered Stokes and anti-Stokes light intensity in a
homogeneous system as a function of the reduced temperature $T/T_c$,
normalized to the scattered intensity at $T_c$ . The inset shows the BCS gap
$\Delta(T)$.}
\label{Slichterfig}
\end{center}
\end{figure}

    To illustrate the effect, we evaluate Eq.\ (\ref{Slichter}) numerically,
with a pairing gap obtained by solving the BCS gap equation as a function of
$T$.  We have chosen $k_F|a|=0.3$ and a frequency shift
$\tilde{\omega}/\epsilon_F=0.0001$, where $\epsilon_F$ is the Fermi energy.
Figure \ref{Slichterfig} shows the temperature dependence of the scattered
light intensity of the Stokes and anti-Stokes lines.  We see that close to
$T_c$ the intensity of light scattered from the superfluid state is
significantly larger than from the normal state.  The intensity for
scattering from the normal state is $\propto T$.  The scattered light from the
superfluid is very large close to $T_c$ due to the density of states effect
described above, while it becomes exponentially suppressed for $T\rightarrow
0$, since the density of quasiparticles available for scattering scales as
$\exp(-\beta\Delta)$.  Due to the large peak in the scattering intensity below
$T_c$, a Stokes--anti-Stokes experiment could clearly reveal the presence of
pairing.  Note that as $\tilde{\omega}$ decreases, the peak below $T _c$
grows.

    To study the effect of the trapping potential, we now examine the
scattered light from a gas in a spherical trap $V_{\rm
pot}(r)=m\omega_Tr^2/2$.  The Cooper pairing is between atoms in time-reversed
angular momentum states, $(l,m)$ and $(l,-m)$, where $l$ is the single
particle orbital angular momentum, and $m$ its component along the incident
laser beam.  We solve the Bogoliubov-de Gennes equations numerically using the
method described in Ref.\ \cite{EPJD}.  With the quasiparticle energies and
wave functions obtained from this calculation, we then compute the scattered
intensity from Eq.\ (\ref{Stokes}), where the quantum number $\eta$ now stands
for $(n,l,m)$, with $n$ the radial quantum number.

\begin{figure}
[htbp]
\begin{center}\vspace{0.0cm}
\rotatebox{0}{\hspace{-0.cm}
\resizebox{7.5cm}{!}
{\includegraphics{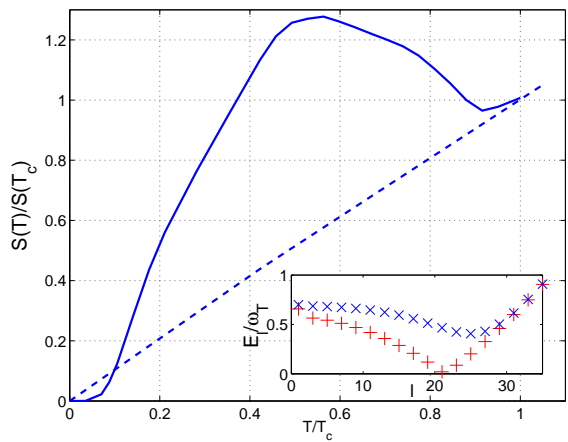}}}
\caption{The scattered Stokes and anti-Stokes light
intensity as a function of $T$,  in units of the scattered intensity
at $T_c$, for a spherically trapped system.  The inset shows the lowest
quasiparticle energies as a function of angular momentum $l$.}
\end{center}
\label{SlichterfigSphere}
\end{figure}

    In Fig.\ 2, we show the calculated intensities, for $1.6\times10^4$
particles trapped with a critical temperature $k_BT_c \simeq 0.09 \epsilon_F$,
$\tilde{\omega}=0.003\epsilon_F$, and $\Delta k=2/l_h$ with
$l_h=(m\omega_T)^{-1/2}$ the trap length.  As in the homogeneous case,
the scattered intensity from the superfluid gas has a large maximum below
$T_c$.  Again, this peak is due to the increased density of states close to
the Fermi level in the superfluid phase for $T$ close to $T_c$.

    The shell structure of normal phase quasiparticle levels is less
pronounced than in the non-interacting case since the Hartree field breaks the
degeneracy of each harmonic oscillator level in the normal
phase~\cite{BruunHeiselberg}.  This introduces a dispersion in the
quasiparticle energies as a function of angular momentum $l$.  The pairing
suppresses this effect bringing the quasiparticles closer in energy, and
yielding a maximum in the density of states close to the Fermi level.  This
effect is shown in the inset in Fig.\ 3 where the
lowest quasiparticle energies are plotted as a function of $l$ for the
superfluid state ($\times$) and the normal state ($+$) for $T/T_c\sim 0.7$.
We see that the dispersion of the energy levels as a function of $l$ is larger
in the normal phase than in the superfluid phase.  Again, it is crucial to
choose a finite momentum shift $\Delta k=2/l_h$ such that the total angular
momentum $l$ of a quasiparticle is not conserved in the light scattering
process.  However for $m$ the angular momentum along the beam axis, a given
quasiparticle with angular quantum numbers $(l,m)$ can scatter to any
quasiparticle state with quantum numbers $(l',m)$ with $l'\neq l$, and the
peak in the density of states close to the Fermi energy in the superfluid
state shows up in the intensity of the scattered light.  For a very small
momentum shift $\Delta k\ll l_h^{-1}$, the scattered signal would, as for the
homogeneous case, from Eq.\ (\ref{Stokes}), be $\propto \sum_\eta
f_\eta(1-f_\eta)$, which does not exhibit any peak below $T_c$ due to pairing.
In contrast to the homogeneous case, the increase in the scattering intensity
due to pairing is smooth at $T_c$ due to the finite size of the spherical
system.  Again, the size of the peak increases with decreasing
$\tilde{\omega}$.

    In summary, we propose the detection of pairing in a two component atomic
Fermi gas by looking at off-resonant light scattering.  The onset of Cooper
pairing yields a large peak in the intensity of the Stokes and anti-Stokes
lines below $T_c$ for both a homogeneous and a trapped gas.  The proposed
effect, which is the light scattering analog of the famous Hebel-Slichter
effect on the nuclear relaxation rate in conventional superconductors, is
unique in that the superfluidity gives an enhanced signal close to $T_c$.  The
interpretation of the results it is not beset by non-trivial mean field energy
shifts.

    We are thankful to J.\ H.\ M\"{u}ller, and S.\ D.\ Gifford for useful
discussions, and to B.\ DeMarco for comments.  This work was supported in part
by NSF Grant PHY00-98353.

\end{document}